\documentclass[twocolumn,english,aps,prl,superscriptaddress]{revtex4-1}
\usepackage[T1]{fontenc}
\usepackage[latin9]{inputenc}
\usepackage{color}
\usepackage{amsmath}
\usepackage{amssymb}
\usepackage{graphicx}
\usepackage{esint}
\usepackage[linktocpage=true,
  colorlinks=true, 
  pdfborder={0 0 0},
  linkcolor=blue,
  citecolor=red,
  filecolor=yellow,
  bookmarks,
  pdftitle={High Tc via spin fluctuations from incipient bands: application to monolayers and intercalates of FeSe},
]{hyperref}

\usepackage{babel}
\begin{document}

\title{High $T_{c}$ via spin fluctuations from incipient bands:\\
 application to monolayers and intercalates of FeSe}

\author{A.~Linscheid}
\affiliation{Department of Physics, University of Florida, Gainesville,FL 32611}
\author{S.~Maiti}
\affiliation{Department of Physics, University of Florida, Gainesville,FL 32611}
\author{Y.~Wang}
\affiliation{Department of Physics and Astronomy, University of Tennessee, Knoxville, Tennessee 37996}
\author{S.~Johnston}
\affiliation{Department of Physics and Astronomy, University of Tennessee, Knoxville, Tennessee 37996}
\author{P.~J.~Hirschfeld}
\affiliation{Department of Physics, University of Florida, Gainesville,FL 32611}

\date{\today}
\begin{abstract}
We investigate superconductivity in a two-band system with an electron- and hole-like band, where one of the bands is away from the Fermi level (or ``incipient''). We argue that the incipient band contributes significantly to spin-fluctuation pairing in the strong coupling limit where the system is close to a magnetic instability, and can lead to a large $T_{c}$. In this case,  $T_{c}$ is limited by a competition between the  frequency range of the coupling (set by an isolated paramagnon) and  the coupling strength itself, such that  a dome-like $T_c$ dependence on the incipient band position is obtained. The coupling of electrons to phonons is found to further enhance $T_{c}$. The results are discussed in the context of experiments on monolayers and intercalates of FeSe.
\end{abstract}
\maketitle
\emph{Introduction} --- Iron based superconductors (FeSC) form the largest family
of unconventional superconductors (SC) known to us. This includes the
stoichiometric,  pressurized,  doped and/or intercalated versions of quasi-2D
layered Fe-pnictogen or Fe-chalcogen compounds of the 1111, 111, 122, and 11
families. These systems collectively host a  variety of superconducting phases
and exhibit a broad range
$T_{c}$'s~\cite{CH_PhysToday2015,Hirschfeld2016,ChubReview2012,WenLiReview2011},
which still lack a consistent explanation. Among these, the FeSe systems
present a curious phenomenology: almost all members related to this family
--alkali/alkali-hydroxy intercalated FeSe~\cite{Qian2011,Niu2015,Zhao2016},
ammonia intercalated
FeSe~\cite{Guo2014,Sedlmaier2014,Burrard-Lucas2013,Guterding2015}-- exhibit
$T_{c}\sim35-45$~K and even reaching $60-70$~K in the case of single unit cell
(UC) thick FeSe grown on 001-SrTiO$_{3}$
(STO)~\cite{Wang2012,Liu2012,He2013,Lee2014}, {110-STO,
and~\cite{Zhang2015,Zhou2015} 001-BaTiO$_3$ (BTO)~\cite{Peng2014}}.
Even bulk FeSe, which has a $T_{c}$ of only 8~K~\cite{Hsu2008},  exhibits
a maximum $T_{c}$ of 36~K~\cite{Medvedev2009} under pressure.

Thus far, the relatively high $T_{c}$'s in these systems have been
correlated with an increase of the $ab$-plane lattice constant in multi
UC FeSe films~\cite{Tan2013,Liu2015} or an increase in the $c$-axis lattice
constant in the intercalates~\cite{Liu2015}. The evolution
from single- to multi-UC films is not smooth~\cite{Liu2014}, but the general
trend is that $T_{c}$ is suppressed when more layers are added. Another correlation, particular to
the 1 UC FeSe on STO (and BTO), is based on the observation of `replica'
bands $\sim100$ meV below the electron and hole bands~\cite{Lee2014}. This is
indicative of a strong forward-focussed ($\boldsymbol{q}=0$) electron-phonon (e-ph) coupling to
a polar phonon mode of the doped STO substrate, which was recently invoked to
explain the high $T_{c}$ of the interfaces
\cite{Lee2014,Rademaker2016,Wang2016,Li2015}.

This, however, does not explain the relatively high $T_{c}$ in the other FeSe
systems without STO phonons, nor why electron doping the FeSC and removing the
hole Fermi pockets seems generally to enhance
$T_{c}$~\cite{Hirschfeld2016,Miyata2015}.  This empirical trend also appears to
directly contradict the spin fluctuation pairing scenario for FeSC, where the
pair scattering by repulsive interactions between
hole and electron Fermi pockets (separated by ${\bf Q}$) leads to strong
pairing; naively, removing the hole pocket should destroy this interaction and
suppress $T_c$ rapidly.  Indeed, while a recent work on incipient band pairing showed
that if pairing exists making use of  electronic states  at the Fermi
level, superconductivity can also be induced in incipient
bands~\cite{Chen2015}, the authors also found that
moving {the hole band} away from the Fermi level always suppresses $T_c$
within the weak coupling framework.

We address these issues in this Letter by noting that an interesting,
and often overlooked, clue lies in the electronic structure of FeSe systems. Although
all the FeSe systems have electron-like Fermi pockets at the $M-$points,
another common aspect is that they have an incipient hole band $~50-100$ meV
below the Fermi level~\cite{Wang2012,Liu2012,He2013,Lee2014,Qian2011}. This
incipient hole band, neglected in nearly all analyses of  SC pairing, is shown
here to play a crucial role in the strong
coupling regime.

{Since the FeSe systems are generally tetragonal and quasi-2D, it suffices to
consider a rather simple model of their electronic structure.  Here, we adopt a  two
band system with a regular electron band and incipient hole band. We then solve
Eliashberg-type equations in which the pairing glue is provided by the
interband  spin-fluctuations (SF), considered in the  strong coupling regime
close to a magnetic instability. We argue that the FeSe systems   are able to
sustain  stronger electronic  correlations   without developing magnetic long
range order due to the suppression of interband scattering by the incipient
electronic states.  }

The proximity to magnetism is well established in these systems. In fact, the
interpretation of the data in Refs. ~\cite{Tan2013} and \cite{Miyata2015}
suggests that FeSe film thickness as low as 3 UC on STO results in a hole band
crossing the fermi level, which is accompanied by spectral features associated
with a magnetic state, while a large $T_c$ is recovered in such thicker films by
electron doping~\cite{Miyata2015}. Further, a first principles study~\cite{Liu2012a}
of 1 UC FeSe/STO found that FeSe would have a very strong spin-density wave (SDW)
without the substrate induced electron doping, again suggesting strong correlations in the system.

{Accounting  for the close proximity to magnetic order and  strong correlations
allowed for by the incipient band,}
we find that the pairing is dominated by a sharp paramagnon
peak in the SF propagator at an energy $\Omega_{p}$ that is induced {near
energies corresponding to the onset of interband transitions}.
Due to the electronic origin of the bosonic peak, its
position controls both the pairing bandwidth and the coupling strength
(unlike a phonon mechanism where the two are largely independent). As the
peak hardens, the interplay between the above two quantities results
in a non-monotonic behavior of $T_{c}$. {Introducing an additional e-ph
coupling further enhances $T_c$}, as was pointed out
in Ref.~\cite{Chen2015}. We now present the systematics of our result.

\emph{Model} --- We take the electron and hole band dispersions
as ($\hbar=1$)
\begin{equation}
\varepsilon_{\boldsymbol{k}}^{{\scriptscriptstyle {\rm h}}}=-\frac{\boldsymbol{k}^{2}}{2m_{{\rm {\scriptscriptstyle h}}}^{\ast}}+E_{{\rm {\scriptscriptstyle h}}},
\qquad\varepsilon_{\boldsymbol{k}}^{{\scriptscriptstyle {\rm e}}}=\frac{(\boldsymbol{k}-\boldsymbol{Q})^{2}}{2m_{{\rm {\scriptscriptstyle e}}}^{\ast}}+E_{{\rm {\scriptscriptstyle e}}}\label{eq:DispersionRelation},
\end{equation}
where $m_{{\rm {\scriptscriptstyle e}}}^{\ast}$ ($m_{{\rm {\scriptscriptstyle h}}}^{\ast}$)
is the effective electron (hole) band mass, $E_{{\rm {\scriptscriptstyle e}}}$
($E_{{\rm {\scriptscriptstyle h}}}$) is the electron (hole) band
extrema measured relative to the chemical potential $\mu$,
and $\boldsymbol{Q}$ is the wave-vector at which SFs are peaked.
The bandwidth is set by requiring the bands to extend
up to $\varLambda_{{\rm {\scriptscriptstyle B}}}$ around their respective
centers in momentum space. We convert all momenta in our plots to energy using
$m^*_{\text{h}}$. In all numerical plots we fix {$\Lambda_B=900$~meV}
and $E_e=-60$ meV, unless otherwise specified.

\begin{figure}
\begin{centering}
\includegraphics{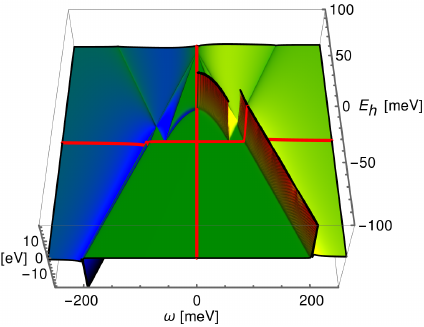}\includegraphics{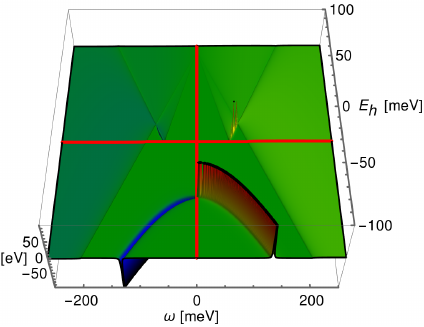}
\par\end{centering}
\caption{(color online) Imaginary part of $V^{{\rm {\scriptscriptstyle SF}}}(\boldsymbol{Q},\omega+{\rm i}\eta)$ for $U=0.325$eV(left) and $U=0.5$eV(right) for a range of $E_{{\rm {\scriptscriptstyle h}}}$.
The Lifshitz transition $E_{{\rm {\scriptscriptstyle h}}}=0$ and the zero energy line $\omega=0$ are highlighted in red.\label{fig:ImVsfRealAxis}}
\end{figure}

In the spin fluctuation framework, the bosonic propagator providing
the pairing glue is the transverse interband spin susceptibility $\chi_{{\rm {\scriptscriptstyle h}}{\scriptscriptstyle {\rm e}}}^{\uparrow\downarrow}(\boldsymbol{q},{\rm i}\nu_{n})$
(we shall henceforth drop the spin indices). Under the usual FLEX-based
approximations \cite{Berk1966,BickersConservingApproximationForStronglyCorrelatedElSystems1989,Graser_etal_2009,EssenbergerSCPairingMediatedBySpinFluctuationsFromFirstPrinciples2015},
the susceptibility acquires a Stoner-like enhancement and the propagator can be modeled as
\begin{eqnarray}
V^{{\rm {\scriptscriptstyle SF}}}(\boldsymbol{q},{\rm i}\nu_{n}) & = & \frac{U^{2}\left(\chi_{{\rm {\scriptscriptstyle h}}{\scriptscriptstyle {\rm e}}}^{0}(\boldsymbol{q},{\rm i}\nu_{n})+\chi_{{\rm {\scriptscriptstyle e}}{\scriptscriptstyle {\rm h}}}^{0}(\boldsymbol{q},{\rm i}\nu_{n})\right)}{1-U\left(\chi_{{\rm {\scriptscriptstyle h}}{\scriptscriptstyle {\rm e}}}^{0}(\boldsymbol{q},{\rm i}\nu_{n})+\chi_{{\rm {\scriptscriptstyle e}}{\scriptscriptstyle {\rm h}}}^{0}(\boldsymbol{q},{\rm i}\nu_{n})\right)},~~~~\text{where}\nonumber\\
\chi_{{\rm {\scriptscriptstyle h}}{\scriptscriptstyle {\rm e}}}^{0}(\boldsymbol{q},{\rm i}\nu_{n}) & = &- \int\frac{{\rm d}^{2}\boldsymbol{k}}{4\pi^{2}}\frac{f(\varepsilon_{\boldsymbol{k}}^{{\rm {\scriptscriptstyle h}}})-f(\varepsilon_{\boldsymbol{k}+\boldsymbol{q}}^{{\scriptscriptstyle {\rm e}}})}{{\rm i}\nu_{n}+\varepsilon_{\boldsymbol{k}}^{{\rm {\scriptscriptstyle h}}}-\varepsilon_{\boldsymbol{k}+\boldsymbol{q}}^{{\scriptscriptstyle {\rm e}}}}
={\chi_{{\rm {\scriptscriptstyle e}}{\rm {\scriptscriptstyle h}}}^{0}(\boldsymbol{q},-{\rm i}\nu_{n})}
.\nonumber\\\label{eq:LindardFunction}
\end{eqnarray}
Here, $f$ is the Fermi function and $U>0$ is a repulsive point-contact
interaction parameter that has the same scale as those in Hubbard-Hund type
models~\cite{BickersConservingApproximationForStronglyCorrelatedElSystems1989,Graser_etal_2009}.
Note that we are discarding the remaining charge and
spin susceptibilties that arise in the FLEX formalism. We therefore expect to
overestimate $T_c$ in this approach.

In the weak coupling limit ($u\equiv m_{{\rm {\scriptscriptstyle e}}}^{\ast}m_{{\rm {\scriptscriptstyle h}}}^{\ast}U/2\pi(m_{{\rm {\scriptscriptstyle e}}}^{\ast}+m_{{\rm {\scriptscriptstyle h}}}^{\ast})\ll1$),
for the incipient case ($E_{\text{h}}<0$), the propagator $V^{{\rm {\scriptscriptstyle SF}}}$
is nonsingular {and can be treated as a constant}~\cite{Chen2015}. For
$u\sim1$, however, the system hosts a sharp paramagnon peak (a pole at $\Omega=\Omega_{\text{p}}$ in the retarded $V^{{\rm{\scriptscriptstyle SF}}}$ ) {controlled by $E_{{\rm {\scriptscriptstyle h}}}$}. {(see Fig. \ref{fig:ImVsfRealAxis})}. The position of this peak determines not only the
size of the electron-SF coupling but also the `pairing bandwidth'
$\varLambda_{{\rm {\scriptscriptstyle SF}}}$ over which
the pairing interaction acts.

{\emph{An estimate for $T_c$}} --- Before we analyze the solutions
to the Eliashberg equations we offer a qualitative picture of the interplay between the coupling
and the pairing bandwidth $\varLambda_{{\rm {\scriptscriptstyle SF}}}$.
To proceed, we focus on $\boldsymbol{q}=\boldsymbol{Q}$; it will
be shown later that for the incipient case ($E_{{\rm {\scriptscriptstyle h}}}<0$),
the $\boldsymbol{q}$ dependence of $V^{{\rm {\scriptscriptstyle SF}}}$
near $\boldsymbol{Q}$ is relatively weak.
Then, for $E_{{\rm {\scriptscriptstyle h}}}<0, ~\Lambda_{\text{B}}\gg E_{\text{h,e}}$
and $T=0$, we obtain
\begin{eqnarray}
\chi_{{\rm {\scriptscriptstyle h}}{\scriptscriptstyle {\rm e}}}^{0}(\boldsymbol{Q},\varOmega_{{\rm {\scriptscriptstyle p}}}) & = & -\frac{m_{{\rm {\scriptscriptstyle e}}}^{\ast}}{2\pi(r+1)}\ln[\frac{r \vert E_{{\scriptscriptstyle {\rm e}}}\vert+\vert E_{{\rm {\scriptscriptstyle h}}}\vert+\varOmega_{{\rm {\scriptscriptstyle p}}}}{\varLambda_{{\rm {\scriptscriptstyle B}}}(1+r)}],\label{eq:Qform}
\end{eqnarray}
where $r=m_{{\rm {\scriptscriptstyle e}}}^{\ast}/m_{{\rm {\scriptscriptstyle h}}}^{\ast}$.
It is clear from Eq.~(\ref{eq:LindardFunction})
that a magnetic instability sets in for $U=U_{c}\equiv1/\left[\chi_{{\rm {\scriptscriptstyle h}}{\scriptscriptstyle {\rm e}}}^{0}(\boldsymbol{Q},0)+\chi_{{\rm {\scriptscriptstyle e}}{\scriptscriptstyle {\rm h}}}^{0}(\boldsymbol{Q},0)\right]$.
For $U<U_{c}$, we have a paramagnon peak at
\begin{eqnarray}\label{eq:ff}
\varOmega_{{\rm {\scriptscriptstyle p}}}^2&=&(|E_{\text{h}}|+r|E_{\text{e}}|)^2-E_0^2\rightarrow 2E_0(E_{\text{h}}^*-E_{\text{h}})
\end{eqnarray}
as the instability is approached, where the paramagnon softens at $E_{\text{h}}=E_{{\rm {\scriptscriptstyle h}}}^{\ast}\equiv-E_0+r \vert E_{{\scriptscriptstyle {\rm e}}}\vert$ and $E_0\equiv\varLambda_{{\rm {\scriptscriptstyle B}}}(1+r){\rm e}^{-\frac{1}{2u}}$.

\begin{figure}[htbp]
\begin{minipage}[t]{0.5\columnwidth}%
\begin{center}
\includegraphics[width=0.95\textwidth]{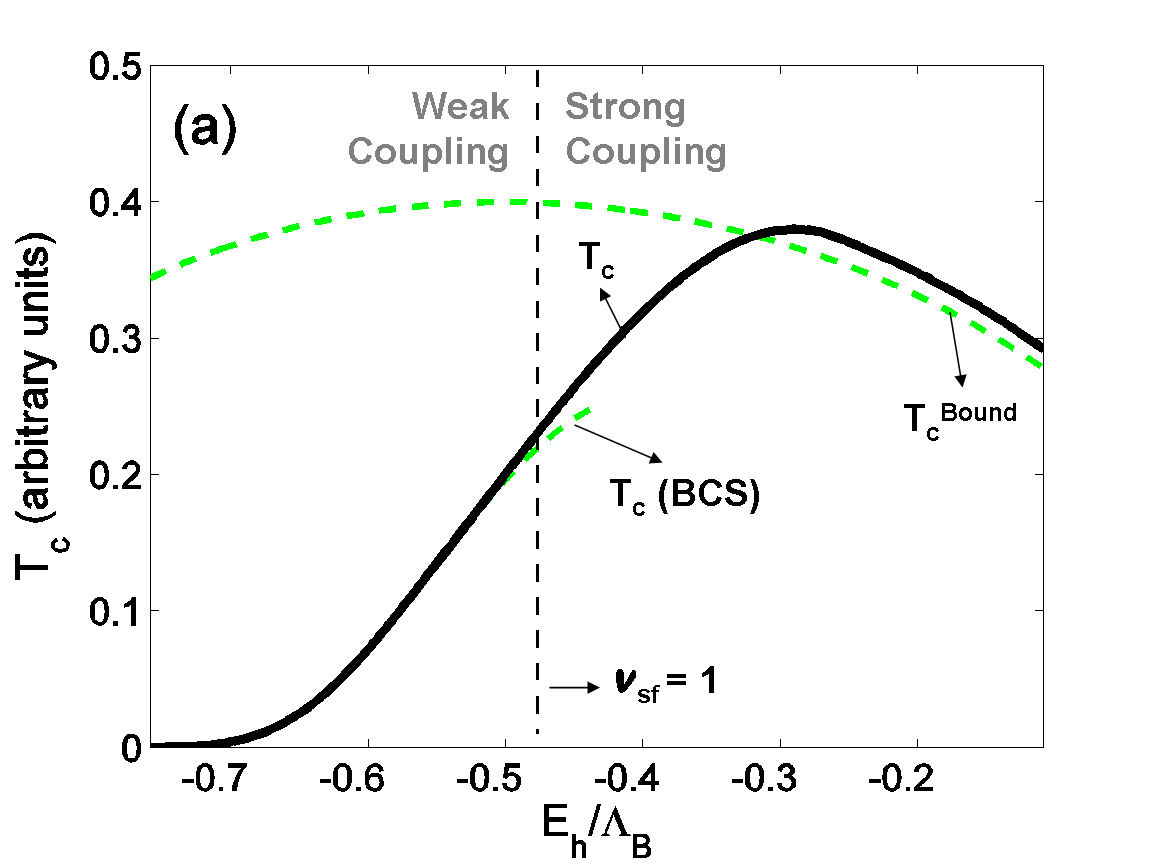}
\par\end{center}%
\end{minipage}\nolinebreak%
\begin{minipage}[t]{0.5\columnwidth}%
\begin{center}
\includegraphics[width=0.95\textwidth]{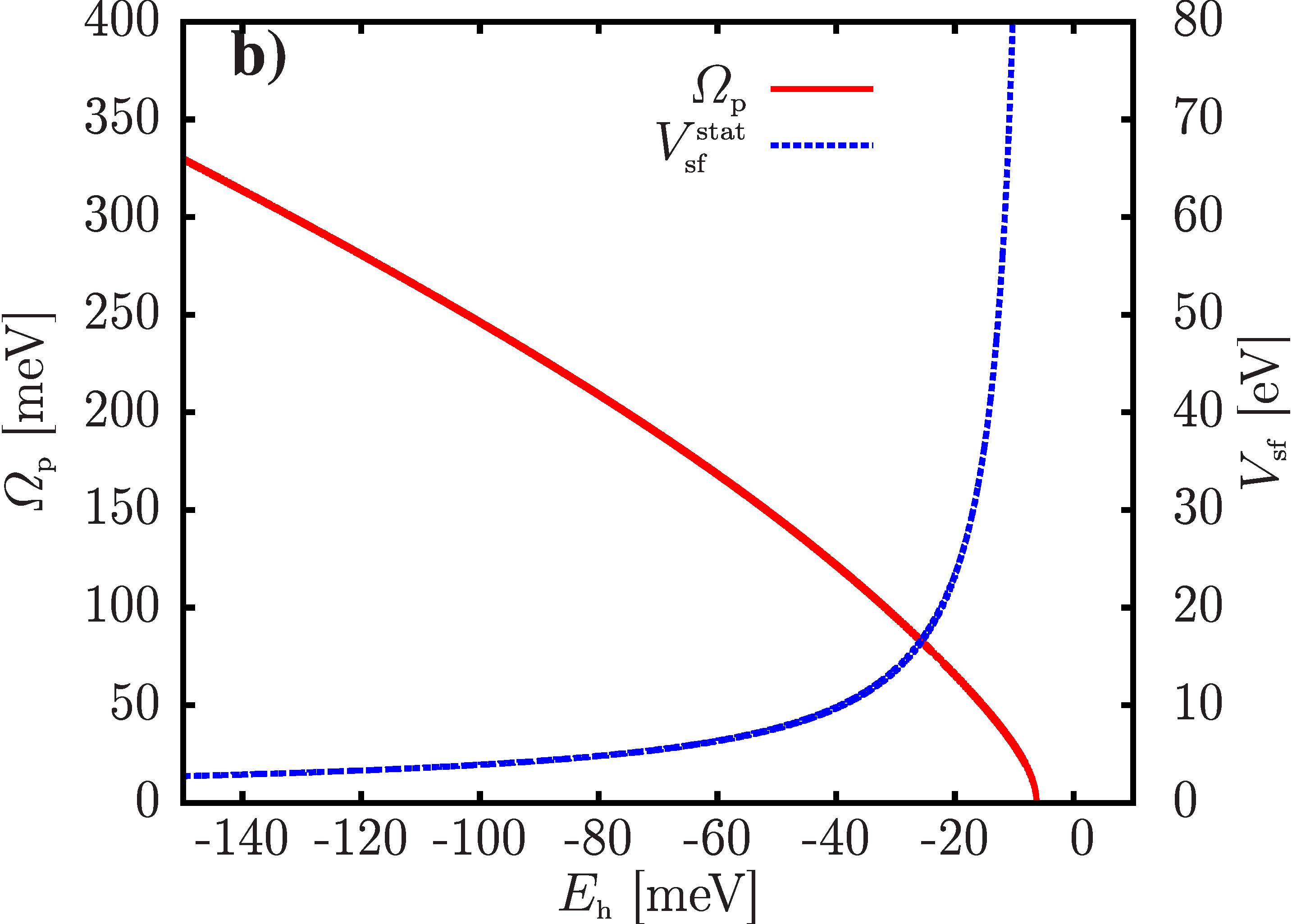}
\par\end{center}%
\end{minipage} \caption{(Color online)
a) Schematic of $T_c$ as $E_{\text{h}}$ is varied. $T_c$ interpolates between
the BCS behavior in the weak coupling regime ($v_{\text{SF}}\ll1$) and the
strong coupling lower bound for $v_{\text{SF}}\gg1$).
b) The
paramagnon and the static propagator $V_{\text{SF}}^{stat}$ ($U=0.45$~eV). Here
$m_{{\rm {\scriptscriptstyle e}}}^{\ast} = 2.6m_e$ and $m_{{\rm
{\scriptscriptstyle h}}}^{\ast}=1.6m_e$.
\label{fig:Tc_Eh}}
\end{figure}

In the usual BCS picture, the dimensionless electron-boson coupling $v_{{\rm {\scriptscriptstyle SF}}}$ is obtained from
the static limit of the propagator multiplied by density of states and the
bandwidth of the pairing interaction $\varLambda_{{\rm {\scriptscriptstyle
SF}}}$ is approximately the position of the bosonic mode $\varOmega_{{\rm
{\scriptscriptstyle p}}}$. Thus,
\begin{eqnarray}
V_{{\rm {\scriptscriptstyle SF}}}^{{\rm {\scriptscriptstyle stat}}}(\boldsymbol{Q}) & = & U\frac{uR}{1-uR}\rightarrow U\frac{E_0}{|E_\text{h}-E_{\text{h}}^*|},\label{eq:ApproxStatCoupl}
\end{eqnarray}
with $R=-2\ln\left[|r
E_{\text{e}}+E_{\text{h}}|/\Lambda_{\text{B}}(1+r)\right]$ and
$v^{\text{e/h}}_{{\rm {\scriptscriptstyle SF}}}=V_{{\rm {\scriptscriptstyle
SF}}}^{{\rm {\scriptscriptstyle stat}}} m^*_{\text{e/h}}/2\pi$. To get an
estimate for $T_c$, we note that as a function of $E_{\text{h}}$, the coupling
$v_{\text{SF}}$ varies from $v_{\text{SF}}\gg1$ near $E_{\text{h}}^*$ to
$v_{\text{SF}}\ll1$ far from $E_{\text{h}}^*$ (see Fig. \ref{fig:Tc_Eh}a). For
$v_{\text{SF}}\ll1$, $T_c$ is given by
$\Lambda_{\text{SF}}\text{exp}[-1/v_{\text{SF}}^2]$; for $v_{\text{SF}}\gg1$,
$T_c$ has the usual Allen-Dynes~\cite{Allen1975} lower bound given by
$\Lambda_{\text{SF}}(v^{\text{e}}_{\text{SF}}v^{\text{h}}_{\text{SF}})^{1/4}$.
Near the instability, the coupling $v_{\text{SF}}$ diverges as
$1/|E_{\text{h}}-E_{\text{h}}^*|$ and the pairing bandwidth
$\Lambda_{\text{SF}}$ vanishes as $\sqrt{|E_{\text{h}}-E_{\text{h}}^*|}$ (as is
clear from Eqs. \ref{eq:ff} and \ref{eq:ApproxStatCoupl}), and the lower bound
\begin{equation}\label{eq:int}
T_c\gtrsim E_0\left(\frac{m^*_{\text{e}}+m^*_{\text{h}}}{\sqrt{m^*_{\text{e}}m^*_{\text{h}}}}\right)^{1/2}
\end{equation}
is then obtained for $T_c$.

If the coupling constant diverges, it is tempting to conclude that
$T_c\rightarrow \infty$ within Eliashberg theory. However, we have demonstrated
that even within an Eliashberg type theory, the $T_c$ remains finite. This is a
hallmark of a strong coupling electronic mechanism for superconductivity.  The
schematic evolution of $T_c$ with $E_{\text{h}}$ based on these considerations
is presented in Fig. \ref{fig:Tc_Eh}b. The suppression of $T_c$ near
$E_{\text{h}}^*$ is a consequence of the dynamics of the pairing interaction,
which leads to a strong mass renormalization of quasiparticles in the
Fermi-liquid state. This aspect is not present in conventional e-ph theories
because the coupling and the boson frequency are usually decoupled.

\begin{figure}
\begin{minipage}[t]{0.33333\columnwidth}%
\begin{center}
\includegraphics[width=0.95\textwidth]{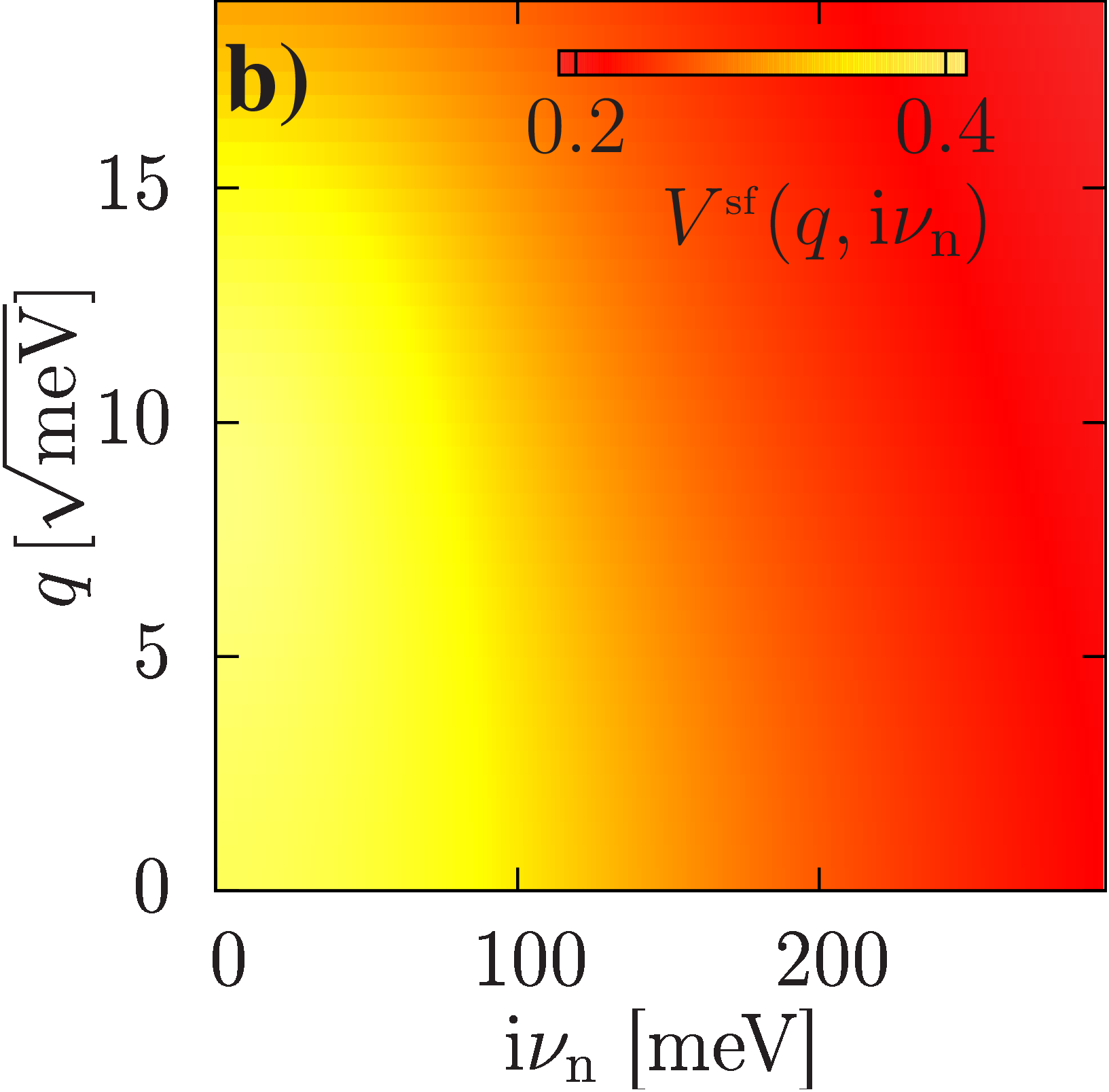}
\par\end{center}%
\end{minipage}\nolinebreak%
\begin{minipage}[t]{0.33333\columnwidth}%
\begin{center}
\includegraphics[width=0.95\textwidth]{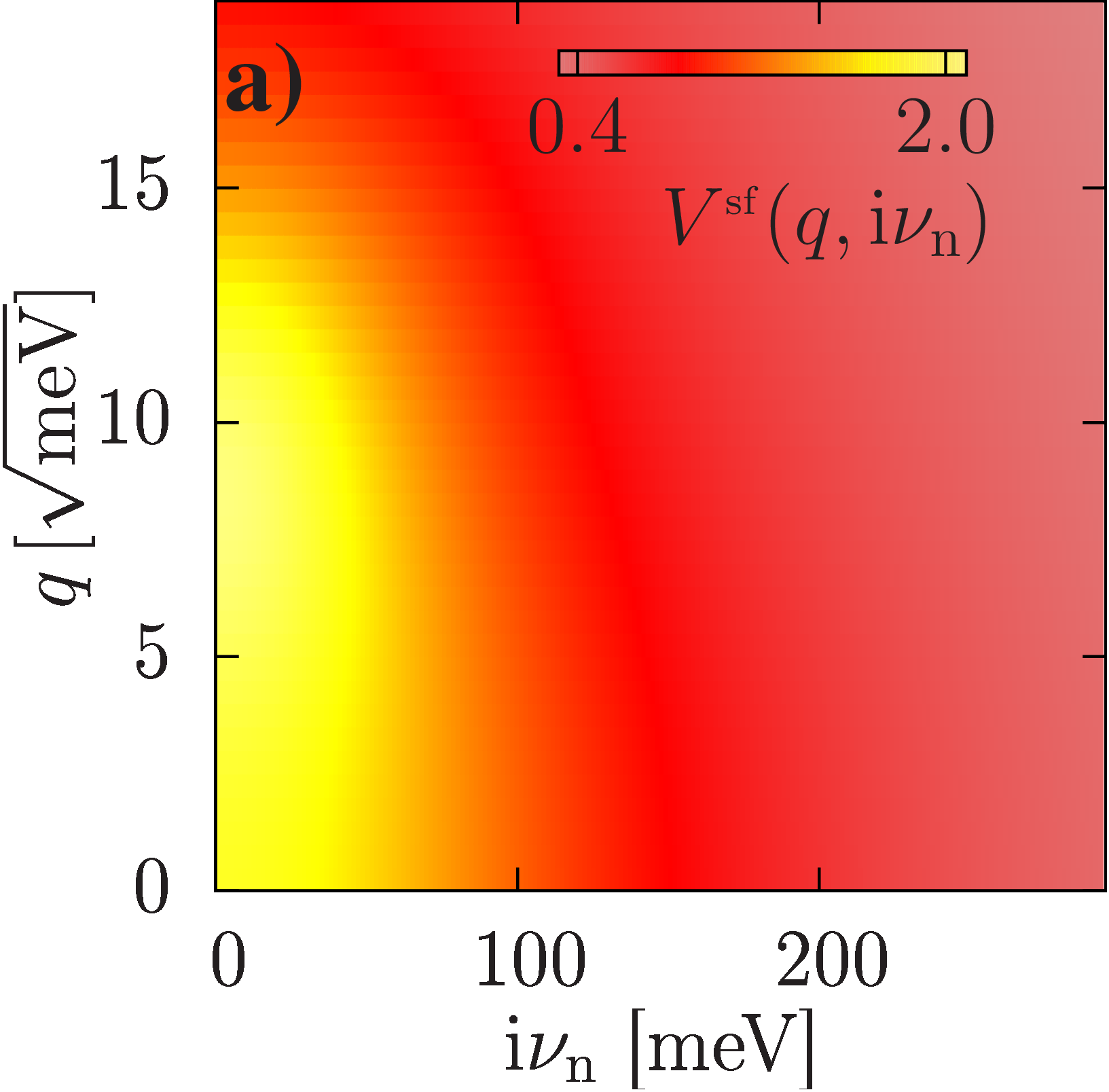}
\par\end{center}%
\end{minipage}\nolinebreak%
\begin{minipage}[t]{0.33333\columnwidth}%
\begin{center}
\includegraphics[width=0.95\textwidth]{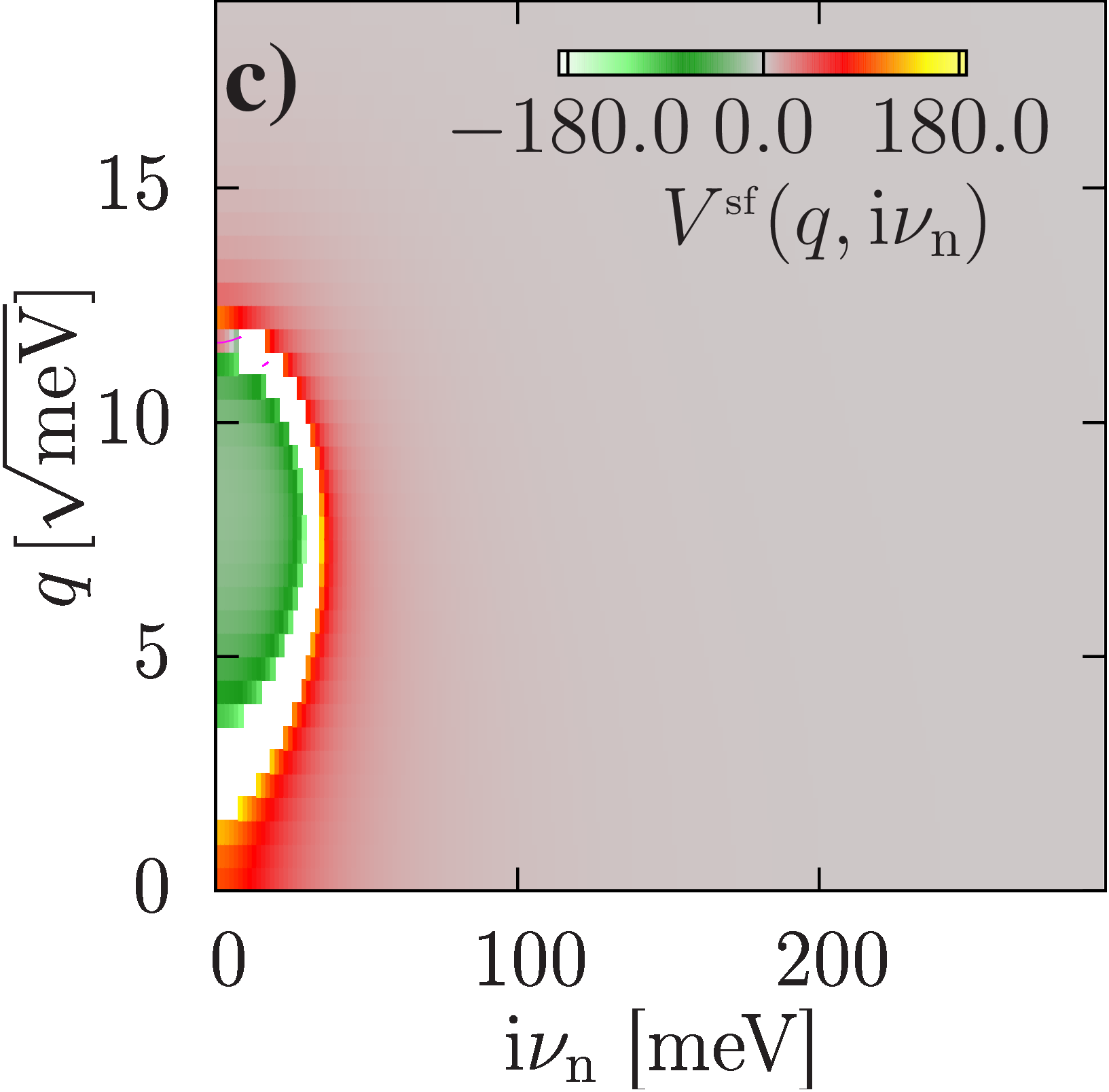}
\par\end{center}%
\end{minipage}\\
\begin{minipage}[t]{0.33333\columnwidth}%
\begin{center}
\includegraphics[width=0.95\textwidth]{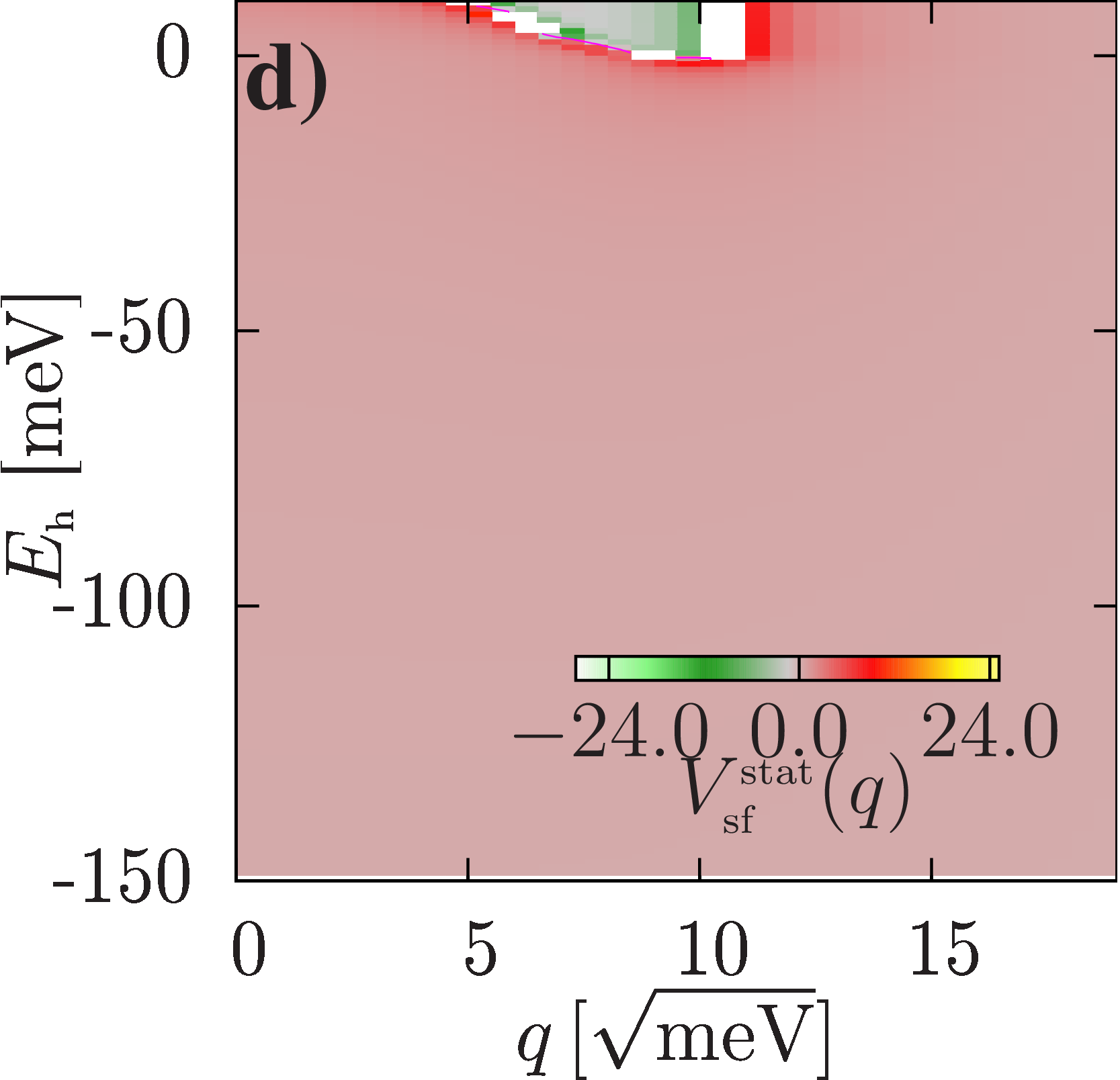}
\par\end{center}%
\end{minipage}\nolinebreak%
\begin{minipage}[t]{0.33333\columnwidth}%
\begin{center}
\includegraphics[width=0.95\textwidth]{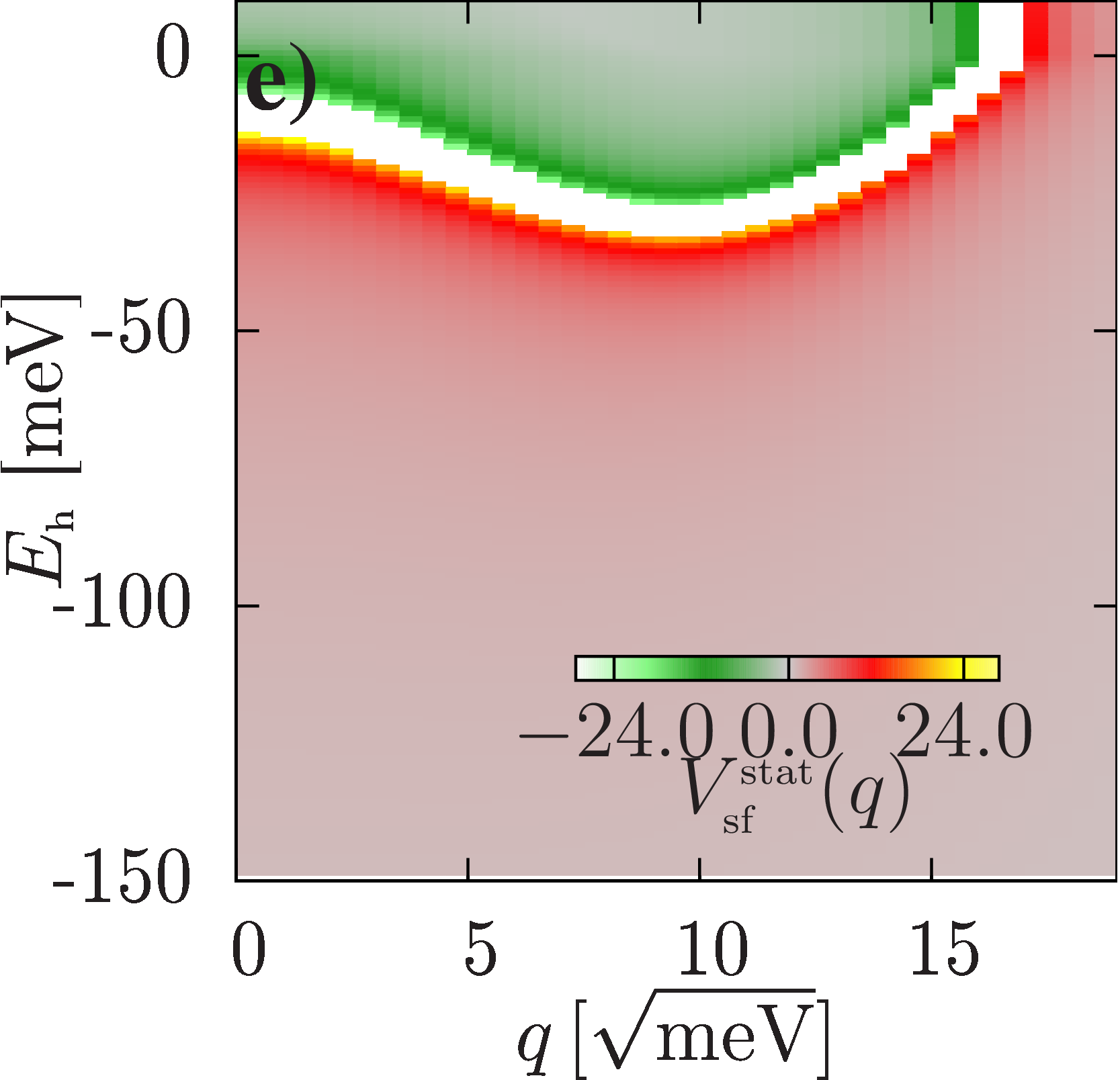}
\par\end{center}%
\end{minipage}\nolinebreak%
\begin{minipage}[t]{0.33333\columnwidth}%
\begin{center}
\includegraphics[width=0.95\textwidth]{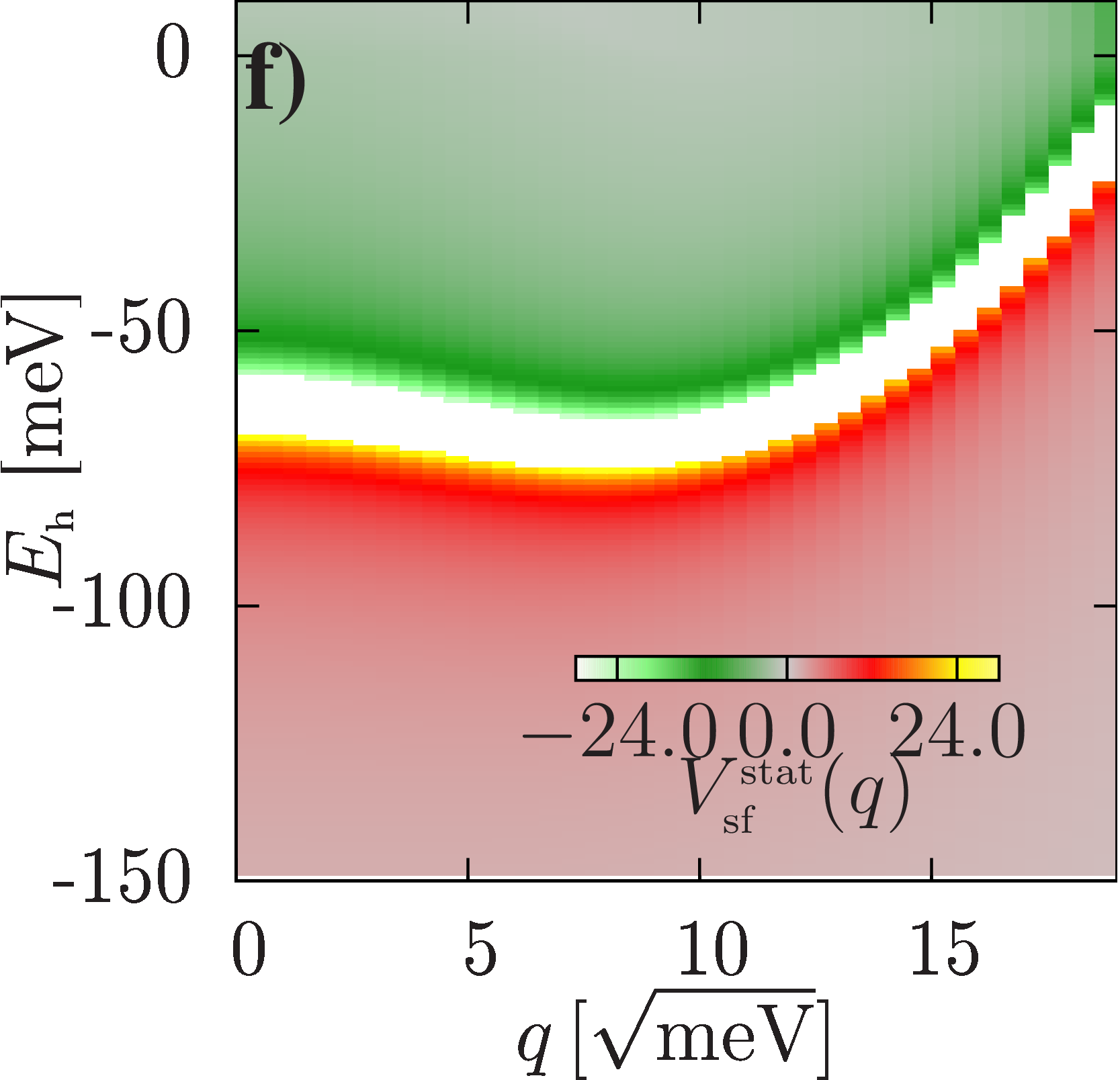}
\par\end{center}%
\end{minipage}

\caption{(color online) Momentum dependence of $V^{{\rm {\scriptscriptstyle SF}}}(q,{\rm i}\nu_{n})$
at $E_{{\rm {\scriptscriptstyle h}}}=-65$ meV for $U=0.4,0.45$ and
$0.51$~eV in a), b) and c) and of the static part $V_{{\rm {\scriptscriptstyle SF}}}^{{\rm {\scriptscriptstyle stat}}}(q)$
for $U=0.4,0.45$ and $0.51$~eV in d), e) and f) as a function Matsubara
frequency and doping $E_{{\rm {\scriptscriptstyle h}}}$, respectively.
$q$ measures the deviation form $\boldsymbol{Q}$.
\label{fig:MomentumDependence} }
\end{figure}

\emph{Incipient Eliashberg equations} ---
To accurately describe the region close to the instability $E_{{\rm {\scriptscriptstyle h}}}^{\ast}$,
we need to turn to a description in terms of the Eliashberg equations. These can be greatly simplified if the effective
interaction $V^{\text{SF}}$ does not  depend strongly on momentum transfer $\boldsymbol{q}$. In Fig.~\ref{fig:MomentumDependence}
we investigate this dependence around $\boldsymbol{Q}$, defining $\boldsymbol{\tilde{q}}=\boldsymbol{Q}+\boldsymbol{q}$. The Stoner-like enhancement leads to a strong variation of $V^{{\rm {\scriptscriptstyle SF}}}(\boldsymbol{q},{\rm i}\nu_{n}= 0)$ close to the instability, as can be seen in Figs.~\ref{fig:MomentumDependence}a) - \ref{fig:MomentumDependence}c).
In Fig.~\ref{fig:MomentumDependence}d), \ref{fig:MomentumDependence}e) and \ref{fig:MomentumDependence}f) we show $V^{{\rm {\scriptscriptstyle
SF}}}(\boldsymbol{q},0)$ as a function of $E_{{\rm {\scriptscriptstyle h}}}$
and the momentum deviation $q$ for the interaction parameters $U=0.4$, $0.45$, and
$0.51$ eV, respectively. The regions shown in green are beyond the instability
and correspond to a magnetic ground state. Thus, in Fig. \ref{fig:MomentumDependence}c),
the magnetic instability occurs at a finite $q$ leading to an incommensurate
SDW ground state. As is evident in Fig. \ref{fig:MomentumDependence}, the momentum
dependence is only important very close to the instability and is rather
featureless when the hole band goes deeper below the Fermi level; we conclude that
the momentum dependence can be neglected in the Eliashberg equations in this
region of parameter space.

\begin{figure*}
\begin{minipage}[t]{0.3333\textwidth}%
\begin{center}
\includegraphics[width=0.95\textwidth]{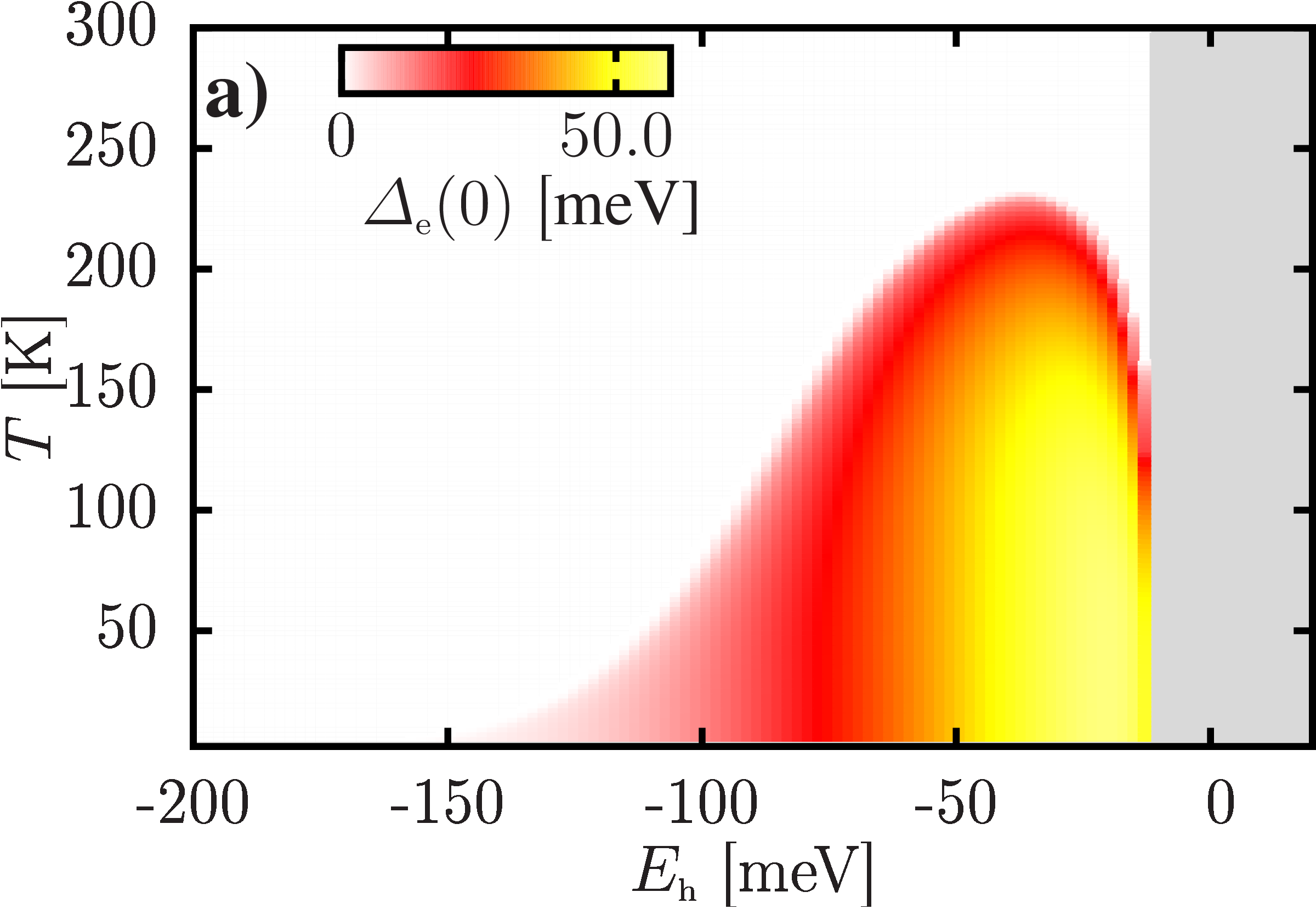}
\par\end{center}%
\end{minipage}\nolinebreak%
\begin{minipage}[t]{0.3333\textwidth}%
\begin{center}
\includegraphics[width=0.95\textwidth]{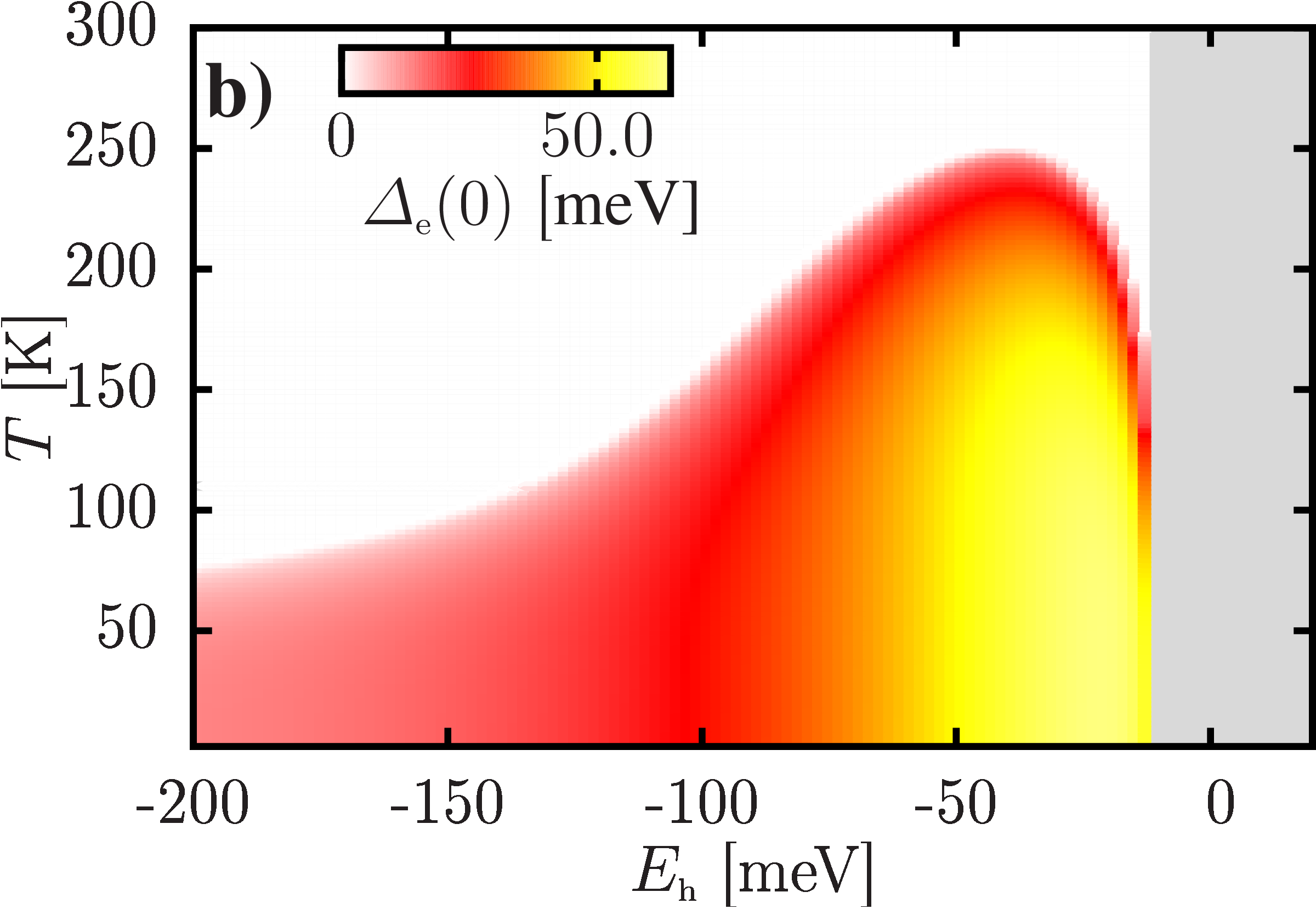}
\par\end{center}%
\end{minipage}\nolinebreak%
\begin{minipage}[t]{0.3333\textwidth}%
\begin{center}
\includegraphics[width=0.95\textwidth]{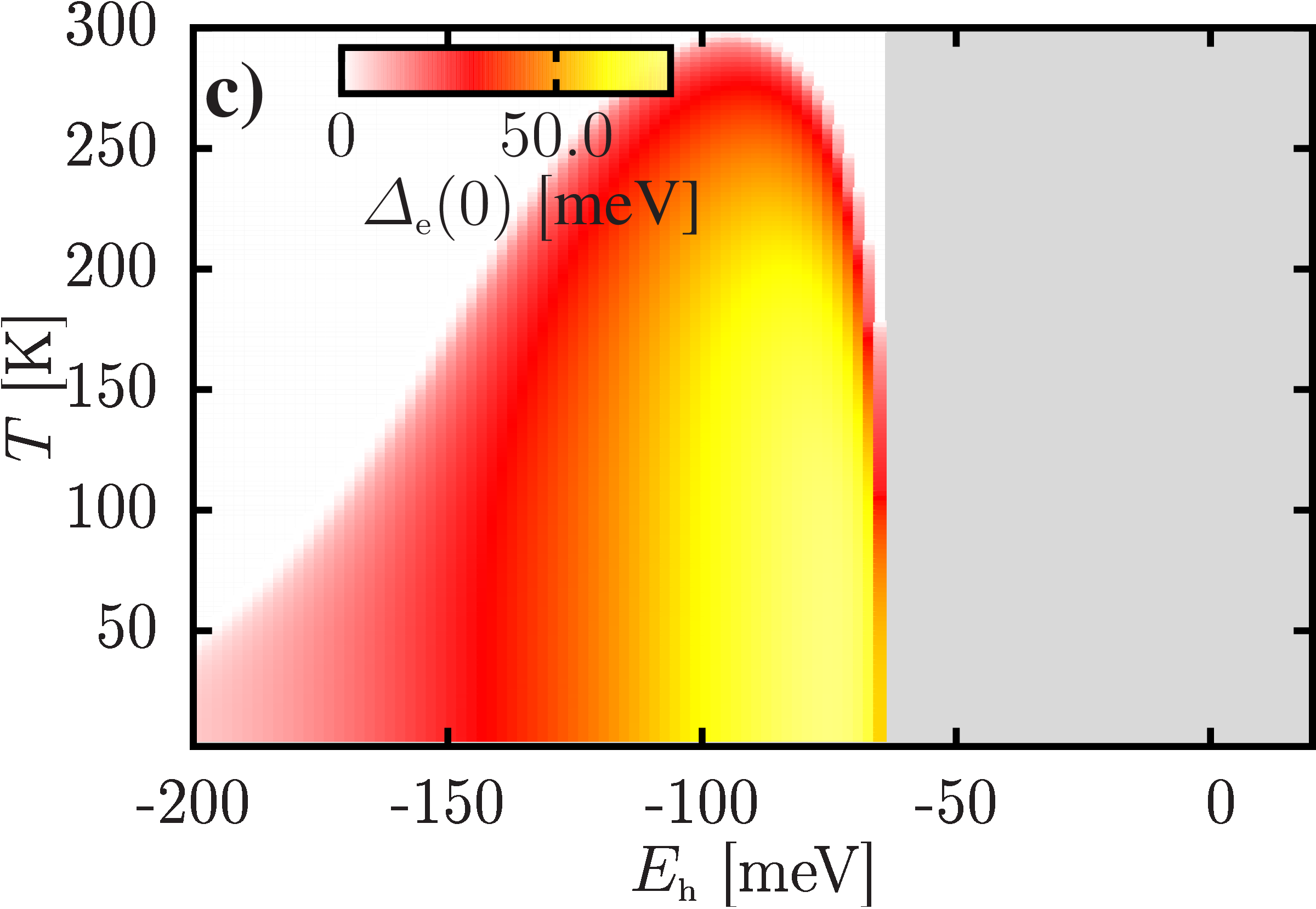}
\par\end{center}%
\end{minipage}\\
\begin{minipage}[t]{0.3333\textwidth}%
\begin{center}
\includegraphics[width=0.9\textwidth]{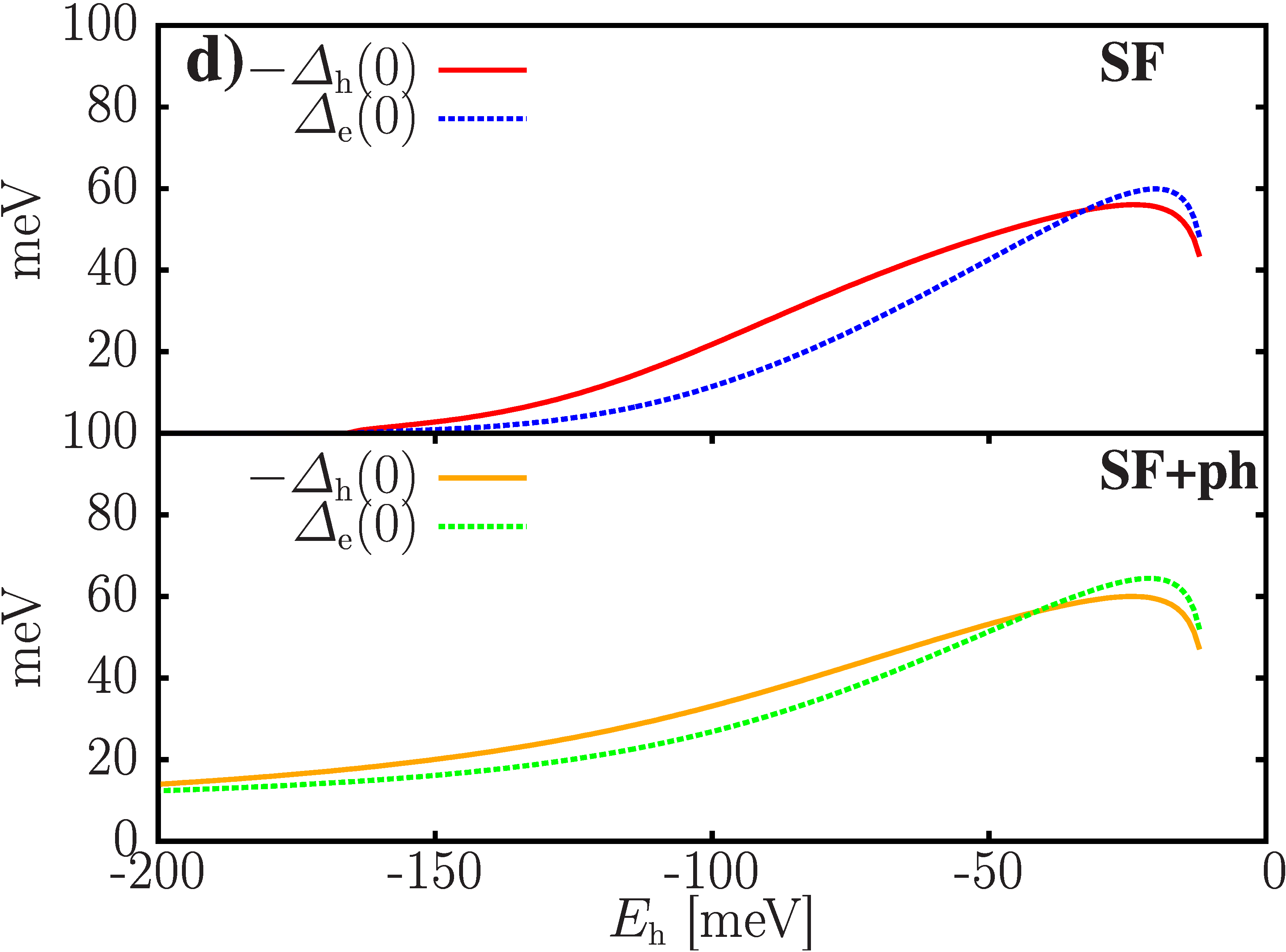}
\par\end{center}%
\end{minipage}\nolinebreak%
\begin{minipage}[t]{0.3333\textwidth}%
\begin{center}
\includegraphics[width=0.9\textwidth]{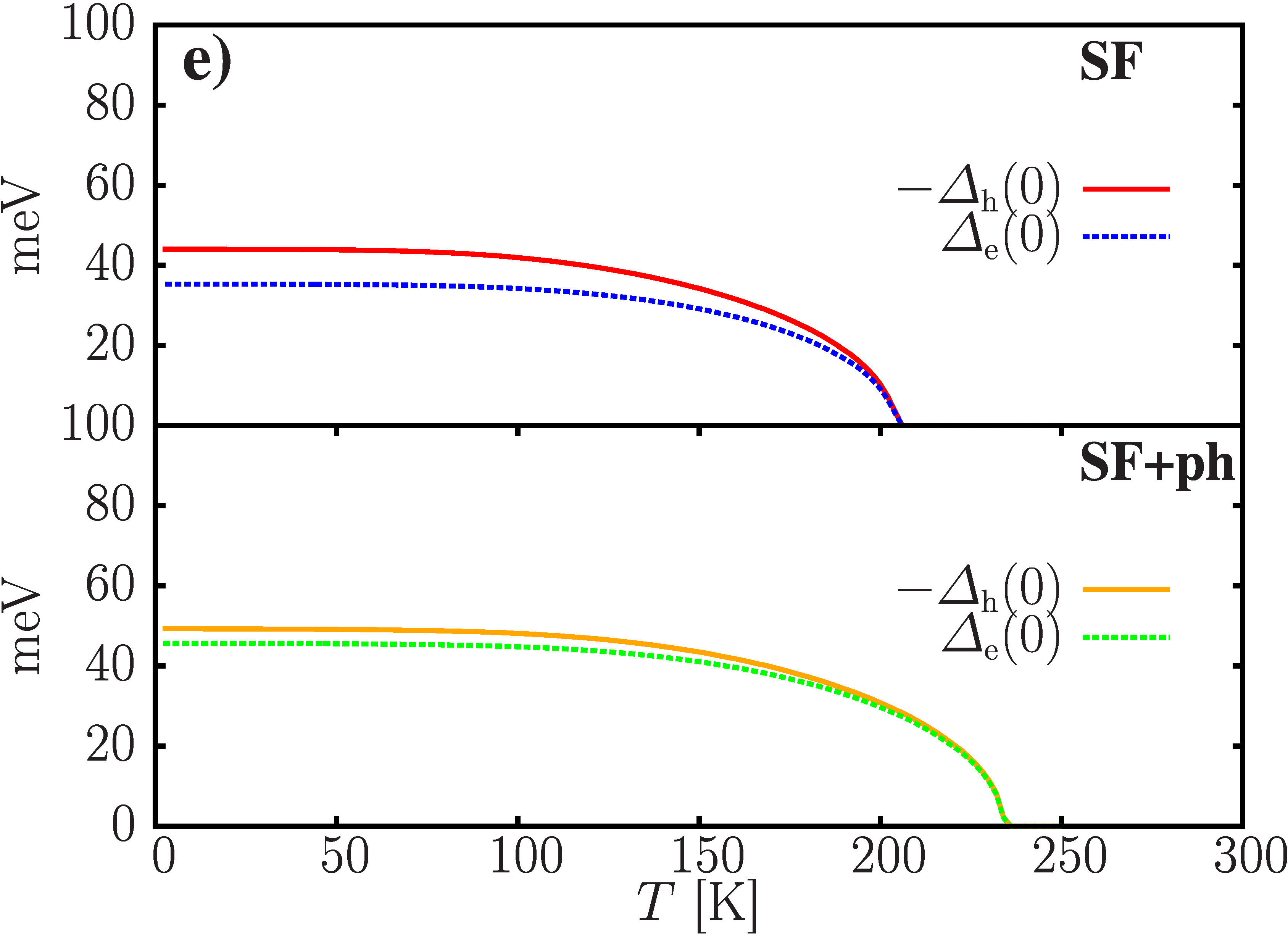}
\par\end{center}%
\end{minipage}\nolinebreak%
\begin{minipage}[t]{0.3333\textwidth}%
\begin{center}
\includegraphics[width=0.9\textwidth]{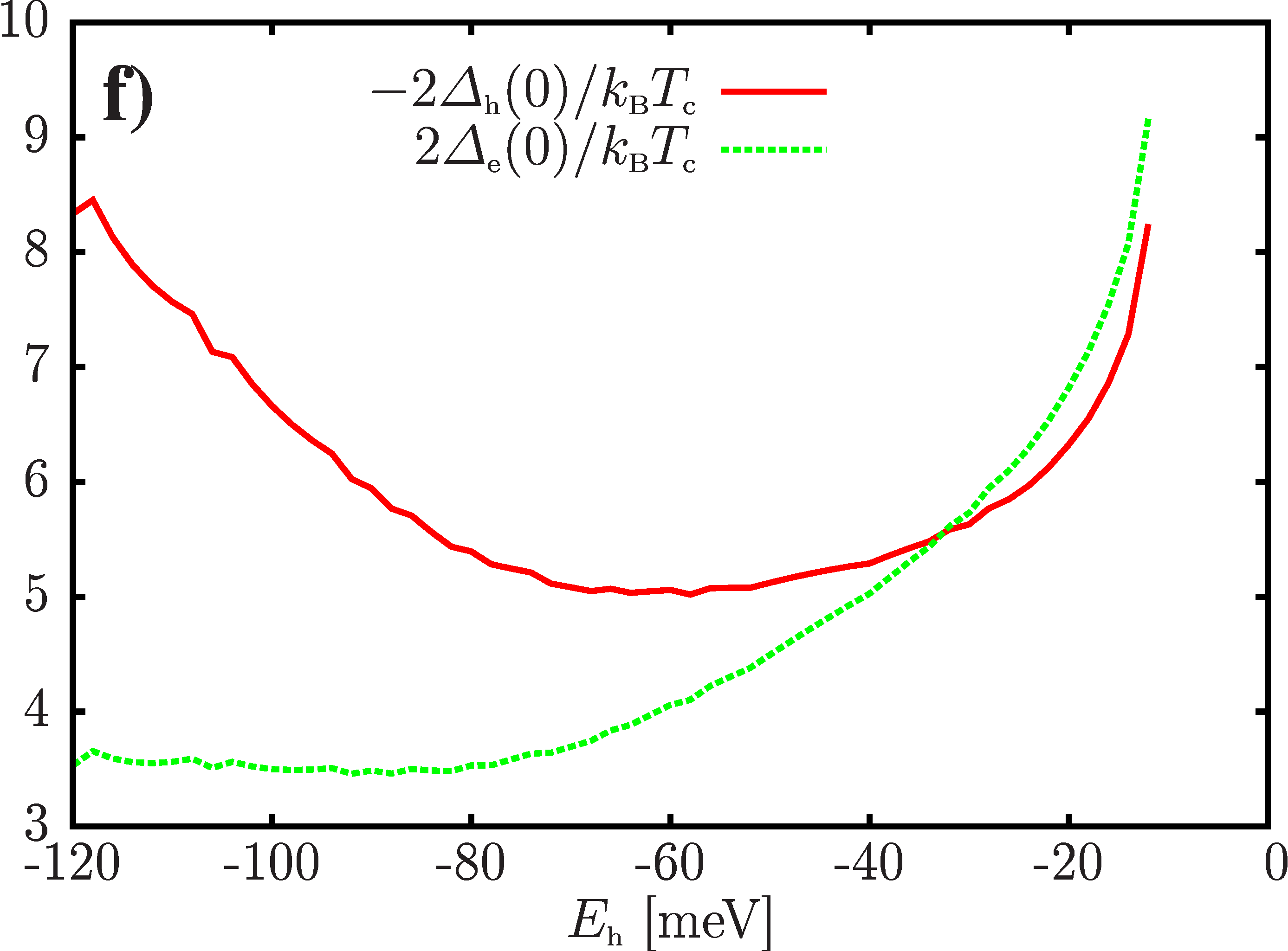}
\par\end{center}%
\end{minipage}

\caption{(color online) Gap on the electron band at the Fermi level as a function
of doping ($E_{{\rm {\scriptscriptstyle h}}}$) and temperature $T$
in a) via spin fluctuations and in b) with an additional intraband
phonon interaction of $\lambda_{{\rm {\scriptscriptstyle ph}}}=0.6$
and $\omega_{{\rm {\scriptscriptstyle ph}}}=100$ meV
Parameters are $U=0.6$ eV and $\varLambda_{{\rm {\scriptscriptstyle B}}}=900~{\rm meV}$. c) same as a) but with $U$ increased to $0.7$~eV. Shaded
areas in a), b) and c) are antiferromagnetic. In d) we plot the two
gaps as a function of doping at $T=2$~K (top) and with the additional
phonon interaction (bottom). e) SF gaps as a function of $T$ (top)
and with the additional phonon interaction (bottom) at $E_{{\scriptscriptstyle {\rm h}}}=-60$ meV.
f) Gap ratios for electron and hole gap vs. $E_{{\scriptscriptstyle {\rm h}}}$.
\label{fig:GapTc}}
\end{figure*}

The multiband isotropic Eliashberg equations in Matsubara axis are (see Supplemental information):
\begin{eqnarray}
Z_{n,i} & = & 1+\frac{T}{\omega_{n}}\sum_{n^{\prime}i^{\prime}}[\lambda_{n-n';i,i^{\prime}}^{{\rm {\scriptscriptstyle ph}}}\xi_{n^{\prime},i^{\prime}}^{{\rm {\scriptscriptstyle ph}}}+\lambda_{n-n';i,i^{\prime}}^{{{\scriptscriptstyle SF}}}\xi_{n^{\prime},i^{\prime}}^{{\rm {\scriptscriptstyle SF}}}]\omega_{n^{\prime}}\nonumber \\
Z_{n,i}\varDelta_{n,i} & = & T\sum_{n^{\prime}i^{\prime}}[\lambda_{n-n';i,i^{\prime}}^{{\rm {\scriptscriptstyle ph}}}\xi_{n^{\prime},i^{\prime}}^{{\rm {\scriptscriptstyle ph}}}-\lambda_{n-n';i,i^{\prime}}^{{\rm {\scriptscriptstyle SF}}}\xi_{n^{\prime},i^{\prime}}^{{\rm {\scriptscriptstyle SF}}}]\varDelta_{n^{\prime},i^{\prime}}.\nonumber\label{eq:IncipientEliashDelta}
\end{eqnarray}
Here, $i,i'\in\{\text{h,e}\}$; $\xi^{\text{P}}_{n,i}=\int_{l_\text{P}}^{h_\text{P}}{\rm d}\varepsilon/D_{n,i}$,
with $D_{n,i}=\varepsilon^{2}+Z_{n,i}^2(\omega_{n}^{2}+\vert\varDelta_{n,i}\vert^{2})$ and $l_\text{P}$, $h_{\text{P}}$ being the cut-offs set by the
mechanism $\text{P}\in\{\text{ph,SF}\}$ (the bandwidth in our calculations);\\ $\lambda_{n-n';i,i^{\prime}}^{{\rm {\scriptscriptstyle SF}}}=N_{i^{\prime}}V^{{\rm {\scriptscriptstyle SF}}}(\boldsymbol{Q},{\rm i}\omega_{n}-{\rm i}\omega_{n^{\prime}})\delta_{i,\bar{i}'}$, where $N_{\text{h,e}}$ are the density of states of the electron and hole bands;
and $\lambda_{n-n';i,i^{\prime}}^{{\rm {\scriptscriptstyle ph}}}=\int{\rm d}\varOmega\,2\varOmega\alpha^{\!2}\! F_{i,i^{\prime}}(\varOmega)/[(\omega_{n}-\omega_{n^{\prime}})^{2}+\varOmega^{2}]$.
We have neglected the single-particle energy renormalization $\chi_{n,i}$.

For the following numerical solution, we choose experimental parameters
for FeSe on ${\rm SrTiO}_{3}$ from Ref.~\cite{Fang2015}: {$m_{{\scriptscriptstyle {\rm h}}}^{\ast}=1.6m_{{\scriptscriptstyle {\rm e}}}$, $m_{{\scriptscriptstyle {\rm e}}}^{\ast}=2.6m_{{\scriptscriptstyle {\rm e}}}$, lattice constant $a=3.9$~\AA, and $E_{\text{e}}=-60$ meV.}
In Fig.~\ref{fig:GapTc}, we show the results from the numerical solution to the
Eliashberg equations. Figure \ref{fig:GapTc}a), shows the gap value $\Delta(i\pi T)$ as a function of
temperature and electron doping ($E_\mathrm{h}$). For the given bandwidth {$\varLambda_{{\rm {\scriptscriptstyle B}}}=900$~meV}, we
observe a maximum $T_{{\rm {\scriptscriptstyle c}}}$
at about the experimental position of the hole band ($E_{{\rm
{\scriptscriptstyle h}}}^{{\rm {\scriptscriptstyle exp}}}=-78$
meV~\cite{Fang2015}) if we assume a reasonable interaction parameter of {$U=0.7$~eV}.

In the panels \ref{fig:GapTc}d) and \ref{fig:GapTc}e), we show the temperature and
doping dependence of the two $s_{\pm}$ gaps, respectively. Note that
the interband nature of the interaction makes the incipient hole
gap larger than the electron band gap. This explains
the counterintuitive trend in Fig.~\ref{fig:GapTc}d) whereby the
electron band gap is suppressed faster by continued electron
doping while the hole band gap reaches its maximum at lower $E_{{\rm
{\scriptscriptstyle h}}}$.

In Fig.~\ref{fig:GapTc}f), we plot the gap/ critical temperature ratio.
For $E_{\rm \scriptscriptstyle h}$ far away from the instability,
we find the BCS value of $3.5$ for $2\Delta_{\rm e}/T_c$,
which increases as $E_h$ moves towards the Fermi level to much larger values of about 9,
 reflecting the
the strong coupling behavior near the magnetic instability.
2$\Delta_{\rm h}/T_c$ has a similar enhanced behavior close to the instability, but
is also much larger than the electron band ratio far from the instability, as discussed
in Ref. \onlinecite{Chen2015}.

\emph{The overestimation of $T_{\rm {\scriptscriptstyle c}}$} ---
Our model produces possible $T_c$ values well above 200~K.
Note, however, that we have neglected the intra-band component of the respulsive
interactions that will reduce this estimate. This is typically
captured by the standard $\mu^*$ approximation; however,
while one can easily show that an analogous RPA treatement leads to a Coulomb
repulsion screened by the electron-band charge suscepbility, the use of the
standard $\mu^*$ approximation is questionable due to the shallowness of the
electron band. A more accurate calculation of this contribution requires
momentum resolution, which would increase the
complexity of our model. Thus, we leave this for future work.
Finally, $T_c$ is also likely to be suppressed in the real system
due to increased fluctations in 2D that are not captured by the
Eliashberg formalism.

\emph{Role of phonons} ---
Since the momentum dependence of the interactions is neglected here,
we cannot account for the forward scattering nature of the
e-ph interaction in the 1UC FeSe/STO. We note, however, that it will lead to a
purely intraband phonon coupling. In this spirit, we add in
Fig.~\ref{fig:GapTc}b) an intra-band phonon interaction of with total coupling
$\lambda_{{\rm {\scriptscriptstyle ph}}}=0.6$, $\Omega_{{\rm {\scriptscriptstyle ph}}} = 100$ meV, and
$\alpha^2F_{i,i^\prime}(\Omega) = \frac{\lambda_{\text{ph}}\Omega_{\text{ph}}}{2}\delta_{i,i^\prime}
\delta(\Omega-\Omega_{{\rm {\scriptscriptstyle ph}}})$, which increases
$T_{{\rm {\scriptscriptstyle c}}}^{{\rm {\scriptscriptstyle max}}}$
to 250~K at the optimal  $E_{\scriptscriptstyle \rm{h}}$ {and makes superconductivity persist to much lower $E_{\scriptscriptstyle \rm{h}}$}. We also note that the $T_c$ increase due to the phonons will likely be larger in a more accurate treatment of the forward scattering and/or the intraband replusion. In the latter case there will a cancellation of the replusive intraband interaction, that will make the influence of the attractive phonon interaction more prominent.

\emph{Conclusions} ---
We have shown that when a magnetic instability nearly coincides with a band
moving below the Fermi level, this so-called incipient band can play an
important role in pairing.  Within a simple two-band model, $T_c$ in such a
system was found to have a dome-like behavior due to the competition between the
coupling strength and spin fluctuation bandwidth. Both of these are controlled
by the paramagnon peak in the SF spectrum, which is present for systems
with strong correlations $(u\sim1)$. For weakly correlated systems ($u\ll1$),
there is no such peak and one recovers the results of Ref \cite{Chen2015}.
With realistic values for the parameters, we find significant optimal values of $T_c$,
even in the absence of Fermi surface-based interactions. In
particular, for values relevant for FeSe/STO (we expect those of the
interaclates to be similar), we find that the maximum $T_{{\rm
{\scriptscriptstyle c}}}$ is clearly capable of explaining the high critical
temperatures observed in these systems. Including an additional phonon coupling
in the energy range of the suspected oxygen modes of STO observed in
Ref.~\cite{Lee2014} further enhances the critical temperature.

\emph{Acknowledgements} --- The authors are grateful for useful discussions with T. Maier, V. Mishra, and D.J. Scalapino. PJH and AL were supported by DE-FG02-05ER46236. S. J. and Y. W. are partially funded by the University of Tennessee's Science Alliance Joint Directed Research and Development (JDRD) program, a collaboration with Oak Ridge National Laboratory.
\bibliographystyle{apsrev4-1}
\bibliography{references}

\end{document}